\documentclass[aps,twocolumn,nofootinbib]{revtex4}

\usepackage{graphicx}
\usepackage{dcolumn}
\usepackage{bm}

\usepackage{amssymb,amsfonts}
\usepackage{epsfig}
\usepackage{epstopdf}
\usepackage[all]{xy}
\usepackage{amsthm}
\usepackage{dcolumn}
\usepackage{hyperref}
\usepackage{url}
\usepackage{dsfont}

\newcommand{\be}{\begin{equation}}
\newcommand{\ee}{\end{equation}}
\newcommand{\bea}{\begin{eqnarray}}

\newcommand{\eea}{\end{eqnarray}}

\def\inbar{\,\vrule height1.5ex width.4pt depth0pt}
\def\IR{\relax{\rm I\kern-.18em R}}
\def\IC{\relax\hbox{$\inbar\kern-.3em{\rm C}$}}

\begin{document}

\title{A small non-vanishing cosmological constant from\\the Krein-Gupta-Bleuler vacuum}

\author{Hamed Pejhan$^1$} \author{Kazuharu Bamba$^2$\footnote{bamba@sss.fukushima-u.ac.jp}}\author{Mohammad Enayati$^3$\footnote{Enayati@iauctb.ac.ir}}\author{Surena Rahbardehghan$^3$\footnote{s.rahbardehghan@iauctb.ac.ir}}
\affiliation{$^1$ Institute for Advanced Physics and Mathematics, Zhejiang University of Technology, Hangzhou 310032, China}
\affiliation{$^2$ Division of Human Support System, Faculty of Symbiotic Systems Science, Fukushima University, Fukushima 960-1296, Japan}
\affiliation{$^3$ Department of Physics, Science and Research Branch, Azad University, Tehran 1477893855, Iran}

\begin{abstract}
We point out a potential relevance between the Krein-Gupta-Bleuler (KGB) vacuum leading to a fully covariant quantum field theory for gravity in de Sitter (dS) spacetime and the observable smallness of the cosmological constant. This may provide a formulation of linear quantum gravity in a framework amenable to developing a more complete theory determining the value of the cosmological constant.
\end{abstract}
\maketitle

Recent observational data \cite{Balbi} reveal that the cosmic expansion is accelerating and point towards a small but non-vanishing positive cosmological constant. This suggests that the Universe might presently be in a dS phase well described by the following conformally flat metric
\begin{eqnarray}\label{metric}
ds^2= dt^2 - \Omega^2(t) \; d\textbf{r}^2= \Omega^2(\tau) (d\tau^2 - \; d\textbf{r}^2),
\end{eqnarray}
in which the scale factor $\Omega(t) = H^{-1}\cosh Ht$ ($H$ is the Hubble parameter), $d\textbf{r}^2$ is the metric on the three-spheres, $t$ and $\tau$ are the proper and the conformal time, respectively.

Inserting the dS metric (\ref{metric}) into the semiclassical Einstein equation with positive cosmological constant $\lambda$,
\begin{eqnarray}\label{EFE}
{\cal{G}}_{\mu\nu} + \lambda g_{\mu\nu}= 8\pi G\langle \hat{T}_{\mu\nu}\rangle,
\end{eqnarray}
and with respect to the fact that the mean value of the energy-momentum tensor $\langle \hat{T}_{\mu\nu}\rangle$ is expected to be proportional to vacuum energy density $\rho g_{\mu\nu}$, the ordinary value for $\rho$ coming from the standard model of particle physics leads to a drastically different cosmology than what we observe today; our Universe becomes dS spacetime with a radius of curvature $\sim mm$, while the most obvious interpretation of recent astronomical data reveals an enormously larger value for the curvature radius of our Universe, $\sim 10^{28}cm$. This is the well-known cosmological constant problem \cite{CCP} (for different approaches to this problem, see, for instance, \cite{Arkani-Hamed,Yokoyama,Carroll}).

This problem mainly seems to be the question of quantum gravity, and we will not attempt to give an answer to it here. Indeed, we need a lot more practical experience with quantum gravity before we can hope to realize such a fundamental issue. Our aim in the present work, instead, will be to point out a potential relevance between the observable smallness of the cosmological constant and a choice of vacuum in the dS gravitational background of our expanding Universe, now known as the KGB vacuum. This vacuum, based on a new representation of the canonical commutation relations, was recently proposed as an alternative to the dS natural vacuum state (the Bunch-Davies state) that yields a fully covariant and coordinate-independent quantization (the KGB quantization) of linearized gravity in dS space \cite{BambaI,BambaII,II,I,Garidi2005,Gazeau1415,de Bievre6230}. [Due to the lack of the natural dS-invariant vacuum state for free gravitons, the fact that is now widely accepted in the physics community (see, for instance, \cite{BambaI,Miao0, RWoodard1430020}), the usual canonical quantization seems to break down for field theory of dS quantum gravity.]

Let us begin by summarizing the general properties of the KGB quantization method, while dS spacetime with the metric (\ref{metric}) covering the whole dS manifold is considered as the classical background. First, the KGB method is a canonical quantization scheme of the Gupta-Bleuler type in the sense that one should distinguish the space based upon which the field operator, the vacuum state and finally the observables are constructed, called the total space of states, from the subspace of physical states based upon which the mean values of the observables are calculated. The total space is the complete, non-degenerate and fully invariant\footnote{The `\emph{full invariance}' here refers to the invariance under the dS isometry group, $SO_0(1,4)$, and also all other possible (including conformal and gauge) invariances of the classical field.} inner-product space generated by all positive and negative frequency solutions (with respect to the conformal time) of the field equation which admits the space of physical states as a closed subspace. This would be a Krein space $=$Hilbert$\oplus$anti-Hilbert \cite{Mintchev} when equipped with the Klein-Gordon inner product. Here, it is convenient to precise the above statements by giving a simple example, a free dS scalar field. Dropping the positivity requirement in the usual rules of canonically quantizing, the KGB quantization would be performed as follows
\begin{eqnarray}\label{field}
\hat{\phi}(x)= \sum_n [ \phi_n(x) \hat{a}_n + \phi_n^{\ast}(x) \hat{a}_n^{\dagger} ] + [ \phi_n^{\ast}(x) \hat{b}_n + \phi_n(x) \hat {b}_n^{\dagger} ],\;\;
\end{eqnarray}
where $[ \hat{a}_n^\dagger , \hat{a}_m ] = \delta_{nm}$ and $[ \hat{b}_n^\dagger , \hat {b}_m ]= -\delta_{nm}$, and the set of $\{\phi_n,\phi_n^{\ast}\}$ are, respectively, positive and negative Klein-Gordon norm modes solving the field equation of motion based upon which the corresponding total space is generated. They generate, respectively, the Hilbert and the anti-Hilbert spaces. The KGB Fock vacuum, or simply the KGB vacuum, then is determined by $\hat{a}_n |0\rangle =0$ and $\hat{b}_n |0\rangle =0$.

Second, the KGB quantization method is indeed a coordinate-independent construction in which the KGB field operator and the associated vacuum state depend \emph{only} on the total space \cite{Garidi2005}. On one hand, this shows that the KGB quantization respects the whole dS manifold symmetries which are basic for field dynamics in this space. More exactly, the KGB quantum field is causal and has all covariance properties of the classical field in a strong sense, even when the KGB method is applied to situations where the usual quantization method fails, such as the dS linearized gravity \cite{BambaI,BambaII,II,I} and the dS minimally coupled scalar field \cite{Gazeau1415,de Bievre6230}. [The crucial point about both cases is that the associated total space is indeed the smallest, complete and non-degenerate inner-product space fulfilling all invariance properties of the theory.] On the other hand, this implies the invariance of the field and the vacuum state under Bogoliubov transformations. Of course, this does not mean that Bogoliubov transformations, which are merely a simple change of the space of physical states, are no longer valid in the KGB quantization method. Indeed, in this context, instead of multiplicity of vacua, we face various possibility for the physical state space depending on the considered spacetime and also on the observer. In other words, the usual ambiguity about vacua is not suppressed but displaced (see the detailed discussion in \cite{Garidi2005}). In this regard and with respect to the argument given by Wald in his seminal work \cite{Wald}, ``\emph{For a spacetime which is asymptotically stationary in both the past and the future we have two natural choices of vacua, and the $S$-matrix should be a unitary operator between both structures}", the KGB structure once again seems to appear in a very natural way. Because, as mentioned by Wald, unitarity of the $S$-matrix would not be preserved if the two vacua are not equivalent.

Third, despite the appearance of negative norm states in this quantization method, the theory gives a meaningful expression for the mean value of the energy-momentum tensor. To see the point, it is convenient to focus on the KGB scalar field given above. As already mentioned, in this context the mean value of the energy-momentum tensor, as an observable, would be evaluated only with physical states, $|\vec{k}\rangle\equiv|{k}_1^{n_1}...{k}_l^{n_l}\rangle = \frac{1}{\sqrt{n_1!...n_l!}}(a_{k_1}^\dag)^{n_1}...(a_{k_l}^\dag)^{n_l}|0\rangle$. On this basis, one can easily see that, due to cancellations between the positive and negative norm parts of the field, the method features a remarkable automatic renormalization mechanism of $\hat{T}_{\mu\nu}$'s which fulfills the so-called Wald axioms \cite{Gazeau1415}; (i) $\langle \hat{T}_{\mu\nu}\rangle$ is covariant and causal because the field is, (ii) no negative energy can be measured in all physical states on which $\hat {b}_n$ vanishes, $\langle\vec{k}|\hat{T}_{00}|\vec{k}\rangle \geq 0$ (the vacuum energy independent of the curvature is zero), (iii) the commutation relations imply that: $\hat {a}_n\hat{a}_n^\dagger + \hat{a}_n^\dagger \hat {a}_n + \hat {b}_n\hat{b}_n^\dagger + \hat{b}_n^\dagger \hat {b}_n = 2(\hat{a}_n^\dagger \hat {a}_n + \hat{b}_n^\dagger \hat {b}_n)$, which reduces to the usual reordering when it is applied to physical states. This renormalization, however, seems to be very different from the other ones in the sense that the trace anomaly does not appear in the KGB vacuum (the expected value of all components of the energy-momentum tensor vanish in the KGB vacuum). Of course, it is not very surprising because, after all, the trace anomaly would appear only through the conformal invariance breaking within the quantization procedure, while the KGB structure preserves all covariance properties (including conformal covariance) of the classical field in a rather strong sense \cite{de Bievre6230}.

Forth, the method recovers all common results in the Minkowskian limit. To see the point, let us consider, a scalar field in Minkowski spacetime with a symbolized interacting lagrangian density ${\cal{L}} = g^{\mu\nu}\partial_\mu \phi\partial_\nu\phi - m^2\phi - V(\phi)$. Recalling that the KGB Fock space of free scalar field is defined by Hilbert$\oplus$anti-Hilbert, the unitary of the theory then would be assured by considering $V'(\Pi_+\hat\phi\Pi_+)$ instead of $V(\hat\phi)$ in deriving the $S$-matrix, in which $\Pi_+ = \sum_{\{\vec{k}\}} |\vec{k}\rangle\langle\vec{k}|$ is the projection over the associated Hilbert space (${\{\vec{k}\}}$ is the complete set of the positive frequency one-particle eigenstates of the corresponding free Hamiltonian). Applying the same procedure on other canonically quantizable theories, by restricting $V$ to the positive energy modes denoted by $V'$, would simply imply the fact that the so-called radiative corrections in the KGB context are the same as in the conventional one. In this sense, it seems that vacuum effects in the KGB quantization scheme only involve \emph{the interacting vacuum} (see the details in \cite{Garidi2005}).

Having reviewed these remarkable properties and with respect to the purpose of this manuscript, the crucial question that must be addressed is whether the KGB expression for the mean value of the energy-momentum tensor defines a meaningful structure for the semiclassical description of general relativity (see Eq. (\ref{EFE})) compatible with the observations. [The KGB vacuum is invariant under the dS group, therefore, the KGB description of the energy-momentum tensor in a $4$-dimensional dS spacetime would be completely determined by its trace, i.e., $\langle \hat{T}_{\mu\nu}\rangle=\langle \hat{T}_{\rho}^\rho\rangle g_{\mu\nu}/4$, in which $\langle \hat{T}_{\rho}^\rho\rangle$ is equal to the change of the effective action $W$ under the conformal transformations $g_{\mu\nu}(x) \rightarrow \Omega^2(x) g_{\mu\nu}(x)$ by virtue of Euler's law; $\langle \hat{T}_{\rho}^\rho\rangle = -\Omega (-g)^{-1/2} \delta{W}/\delta \Omega$ \cite{Birrell}.] As we saw above, the KGB free-fields contributions to the vacuum expectation value of the energy-momentum tensor, quite insensitive to the curvature, is zero. This statement, however, is not the final sentence for the KGB approach to explain the observations. The real world is indeed dominated by interacting fields as well, and therefore, a precise answer to the above question also needs to involve interacting-fields contributions in our dS background.

According to the standard model, matter is made up of leptons and quarks which are interacting through two basic types of interactions: the strong and the electroweak interactions. We begin by QCD, the only well-understood fundamental strongly-interacting QFT realized in nature. Technically, due to the intrinsic difficulties of QCD and strongly-interacting fields in general, we do not aim to give an explicit calculations of the $\langle \hat{T}_{\rho}^\rho\rangle$ for such interacting theories in dS spacetime, instead we just give a preliminary estimate of the expected order of magnitude of $\langle \hat{T}_{\rho}^\rho\rangle$. To do this, we recall that, although the KGB Fock vacuum differs from the usual one, by applying the unitary condition, the KGB interacting vacuum effects in flat spacetime would be the same as the standard QFT \cite{Garidi2005}. Therefore, the same as the usual QFT, we can utilize the famous result of QCD in Minkowski spacetime to give a preliminary estimate of the expected order of magnitude of $\langle \hat{T}_{\rho}^\rho\rangle$ in dS space as (for an adiabatic approach see \cite{Schutzhold}, and others, \cite{Bjorken,Klinkhamer1,klinkhamer2})
\begin{eqnarray}\label{contribution}
\langle \hat{T}_{\rho}^\rho\rangle \sim H_0\Lambda^3_{\mbox{\tiny{QCD}}} \sim (10^{-3} {\mbox{eV}})^4,
\end{eqnarray}
where $H_0$ is the Hubble parameter in the late Universe. Note that, this result nicely fits the empirical data \cite{Balbi}. Nevertheless, one may argue that to build up a true picture of any estimate, including the KGB one, of the vacuum energy density stored in the cosmological constant today, it is  also imperative to consider the electroweak interactions. To answer this claim, we should emphasize that contrary to QCD, such interacting theories are weakly coupled and therefore their contributions to the vacuum energy, in comparison with strongly-interacting QFT, can be ignored for all imaginable applications in our study (see \cite{Schutzhold}). Besides this, as already mentioned, quite contrary to the usual QFT, the contribution of KGB free-fields to the vacuum energy is exactly zero. Therefore, it seems that the estimated order (\ref{contribution}) should be respected as the first-order approximation of the predicted value in the KGB context.

In conclusion, while there is certainly a tremendous amount of work still to be done, recalling the fact that the first step toward a satisfactory quantum theory of gravity is to formulate linear gravity in a consistent framework of QFT determining the value of the cosmological constant, we would argue that the KGB quantization scheme could very well in its own right be a promising candidate for such a consistent framework. Last but certainly not least, the KGB method is one of very few constructions that has any realistic hope of direct confrontation with minimal requirements (the causality and the full dS-covariance) of quantum fields in dS space and observation, and so is well worth a very careful look.

\section*{Acknowledgements}
This work was partially supported by the JSPS KAKENHI Grant Number JP25800136 and Competitive Research Funds for Fukushima University Faculty (18RI009) (K.B.).

\end{document}